\documentclass[twocolumn,prl,aps]{revtex4}

\usepackage{graphicx}
\usepackage{dcolumn}
\usepackage{bm}

\begin{document}

\title{Ferromagnetic `$\pi$'-junctions}
\author{T. Kontos, M. Aprili, J. Lesueur, F. Gen\^{e}t, B. Stephanidis and R. Boursier}
\affiliation{CSNSM-CNRS, Universit\'{e} Paris-Sud, 91405 Orsay Cedex, France}
\date{\today}

\begin{abstract}
We investigate Josephson coupling through a ferromagnetic thin film using
Superconductor-Insulator-Ferromagnet-Superconductor planar junctions. Damped oscillations of the critical current are
observed as a function of the ferromagnetic layer thickness. We show that they result from the exchange energy gained
or lost by a quasiparticle Andreev-reflected at the ferromagnet-superconductor interface. The critical current cancels
out at the transition from positive ("$0$") to negative ("$\pi$") coupling, in agreement with theoretical calculations.
\end{abstract}
 \maketitle

Current can flow without dissipation through a thin insulating layer or a weak link separating two superconductors.
Because of the quantum character of the superflow a phase difference, $\theta$, between the superconductors appears.
The current-phase relationship of the link is a periodic function of $\theta$. For a tunnel barrier, as first found by
Josephson, \cite{jos62} it is given by $I=I_{c}\sin(\theta)$, where $I_{c}$ is the critical current. In this Letter, we
show that $I_{c}$ can change sign when the superconductors are coupled via a thin ferromagnetic layer. Referring to the
Josephson current-phase relationship, this change corresponds to a $\pi$-phase shift of $\theta$. Thus Josephson
junctions presenting a negative coupling are usually called $\pi$-junctions. We specifically observe the transition
from a positive to a negative coupling or equivalently from $"0"$ to $"\pi"$ junctions by increasing the ferromagnetic
layer thickness.

The possibility of a negative Josephson coupling was first raised by Kulik \cite{kul66}. It is due to the spin-flip of
the electrons forming a Cooper pair when tunneling through an insulator. The spin-flip originates from the electron
exchange interaction with uncorrelated magnetic impurities in the barrier.  This channel, which coexists with direct
tunneling, is suggested to reduce the critical current of the junction. Whether or not spin-flip tunneling can overtake
direct tunneling, leading to a negative critical current, remains uncertain.

More recently Buzdin et al. \cite{buz81} showed that in ballistic Superconductor/Ferromagnet/Superconductor weak links
(S/F/S), $I_{c}$ displays damped oscillations as a function of $d_{F}$, the thickness of F. Close to the critical
temperature of the superconductor an analytical expression of $I_{c}$ is available. They found $I_{c} \propto
\sin\phi_{F}/\phi_{F}$, where $\phi_{F}= 2E_{ex}d_{F}/\hbar v_{F}$. $E_{ex}$ is the exchange energy and $v_{F}$ the
Fermi velocity in F. When $\phi_{F}$ is greater than $\pi$ and smaller than $2 \pi$, $I_{c}$ is negative leading to a
$\pi$-coupling. Physically, this is a consequence of the phase change of the pair function induced in F by the
proximity effect \cite{kon01}. In fact, energy conservation requires that a Cooper pair entering into the ferromagnet
receives a finite momentum, $\Delta p=\hbar v_{F}/2E_{ex}$, from the spin splitting of the up and down bands
\cite{dem97}. By quantum mechanics, $\Delta p$ modifies the phase, $\phi=\Delta p \cdot x$, of the pair wave function
that increases linearly with the distance, $x$, from the S/F interface. If all pair trajectories and spin
configurations are taken into account, the amplitude of the superconducting order parameter in F is given by
$\sin(\phi)/\phi$ \cite{dem97}. When F is coupled to a second superconductor through a weak link, the Josephson
critical current is proportional to the pair amplitude in F and hence to $ \sin\phi_{F}/\phi_{F}$.

\begin{figure}
\includegraphics[width=8.0cm,keepaspectratio=true]{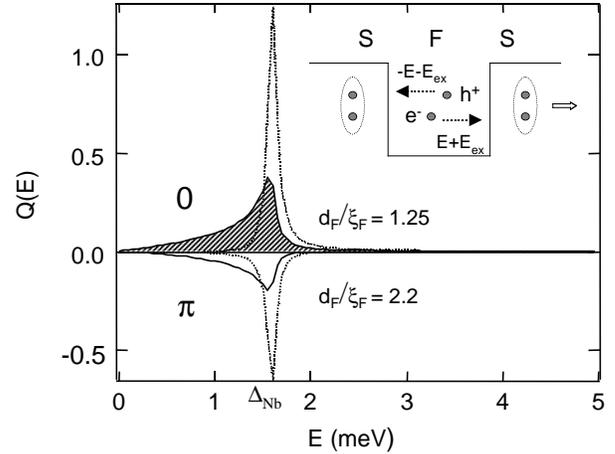}
\caption{Spectral current density, $Q(\epsilon)$, as a function of energy for SFS (full line) and SIFS (dashed line)
Josephson junctions. $Q( \epsilon)$ is calculated from the Usadel equations \protect\onlinecite{yip98} for two
different $d_{F}/ \xi_{F} $ ratios. The "$0$" ($d_{F}/ \xi_{F} $ = 1.8) and "$\pi$" ($d_{F}/ \xi_{F} $ = 3) couplings
result from positive (dashed area) and negative integrals respectively.  The inset illustrates the transfer mechanism
for Cooper pairs through F by Andreev reflections.} \label{fig1}
\end{figure}

Microscopically the transfer of Cooper pairs through the ferromagnet occurs via Andreev reflections \cite{and64}. An
electron-like excitation in F with energy lower than the superconducting energy gap, $\Delta$, cannot enter into the
superconductor. It is reflected at the F/S interface as a hole and it is then reflected back as an electron at the
opposite S/F interface as illustrated in the inset of Fig.\ \ref{fig1} \cite{kul70}. The constructive interference of
the electron-like and hole-like excitations gives rise to Andreev bound states (ABS) in F, which carry the
supercurrent. The same mechanism accounts for the Josephson coupling through a normal metal \cite{kul70}, but in a
ferromagnet the spectrum of the ABS is affected by the spin splitting of the spin bands as the Andreev reflections
reverse the spin of the quasiparticles \cite{kup90}. Therefore the spectral current density, $Q(\epsilon)$ that defines
the current density per bound state in F is a function of the exchange energy, $E_{ex}$. In the dirty limit,
$Q(\epsilon)$ can be obtained by solving \cite{yip98} the Usadel equations \cite{usa70}. The exponential decay of the
order parameter in F on a length scale given by $\xi_{F}=\sqrt{\hbar D/2E_{ex}}$, the coherence length in F, allows use
of the "long junction" approximation \cite{siv72}, in which the thickness of the  ferromagnetic layer, $d_{F}$, is on
the order of $\xi_{F}$ and D is the diffusion constant. When the energy gap is smaller than $E_{ex}$ and the Thouless
energy $E_{Th}=\hbar D/d_{F}^2$, as in the experiment described below, the energy dependence of $Q(\epsilon)$ is
basically fixed by the BCS superconducting density of states. Therefore, the spectral current density is peaked at the
superconducting energy gap. It is either positive or negative for  $E_{ex}< 9\pi ^2 /16 E_{Th}$ (i.e. $d_{F}< 3\pi /4
\xi_{F}$) or $E_{ex}> 9\pi ^2 /16 E_{Th}$ (i.e. $d_{F}> 3\pi /4 \xi_{F}$) respectively as shown in Fig.\ \ref{fig1}.
The critical current $I_{c}$ at zero temperature is obtained by integrating $Q(\epsilon)$; for $d_{F}/\xi_{F}>>1$,
$I_{c}=3.21\Delta / R_{n} \Big[d_{F}/ \xi_{F}\Big] \exp\Big[-d_{F}/\xi_{F} \Big]\sin\Big[\pi/4+ d_{F}/ \xi_{F}\Big]$ as
found by Buzdin \cite{buz91}. Thus, we recover quantitatively the behavior illustrated above.

Experimentally, since the coherence length in F is very small ($\xi_{F} \simeq 30$ \AA, as shown below) the simplest
way to measure the critical current is to fabricate planar junctions where a thin insulating barrier is inserted
between the superconducting counter-electrode and the ferromagnet. The tunnel barrier raises the junction resistance,
$R_{n}$, lowering $I_{c}$ to measurable values (typically 10-100 $\mu A$). In a
Superconductor/Insulator/Ferromagnet/Superconductor (SIFS) Josephson junction the energy spectrum of the Andreev bound
states in F is modified by the insulating layer \cite{zai90}. However, the current spectral density reported in Fig.\
\ref{fig1} shows the same general behavior as observed in SFS junctions, including the change of sign of the Josephson
critical current as a function of the ferromagnetic layer thickness.

Planar $Nb/Al/Al_{2}O_{3}/PdNi/Nb$ junctions were fabricated by e-gun thin film evaporation in a typical base pressure
of $10^{-9}$ Torr, rising to $10^{-8}$ Torr during deposition. The film thickness was monitored during growth to better
than 1 {\AA} by a quartz balance. The junction had the standard four terminal cross-geometry \cite{kon01}, defined
during evaporation by shadow masks. A Si wafer was first covered by 500 {\AA} of SiO. Then, a 1500 {\AA} thick Nb strip
was evaporated and backed by 500 {\AA} of Al. The critical temperature of the Nb strip was 9.2 K and its
residual-resistance-ratio ratio about 4. The Al surface oxidation was performed in a glow discharge during 1 min, and
completed in a 10 Torr oxygen atmosphere during 2 min. A square window of $1mm \times 1mm$ left in a 500 {\AA} thick
SiO layer deposited on the Al defined the junction area. Eight  junctions were aligned on the same Nb/Al  strip. A PdNi
strip (thickness 40-150 {\AA}) perpendicular to the Nb/Al strip was then evaporated on each junction and covered by a
500 {\AA} thick Nb counter-electrode ($T_{c}$=8.7 K). The junction resistance varied from 50 $m\Omega$ to 400
$m\Omega$. The entire process was carried out without breaking the vacuum in the deposition chamber. The Ni
concentration was about 12 $\%$ as checked by Rutherford Backscattering Spectrometry on samples evaporated at the same
time. At this concentration, the PdNi is an itinerant ferromagnet \cite{bei75}, whose the Curie temperature is about
100 K. The junction was polarized by an ac current, the I-V characteristics being monitored on a digital commercial
oscilloscope set in the X-Y mode and directly recorded from the oscilloscope.  Measurements were performed down to 1.5
K; the mimimum detectable value of $I_{c}$ was about 1 $\mu A$. A $\mu$-metal shield reduced the residual magnetic
field on the sample to $10^{-2}$ Gauss.

\begin{figure}
\includegraphics[width=8.0cm,keepaspectratio=true]{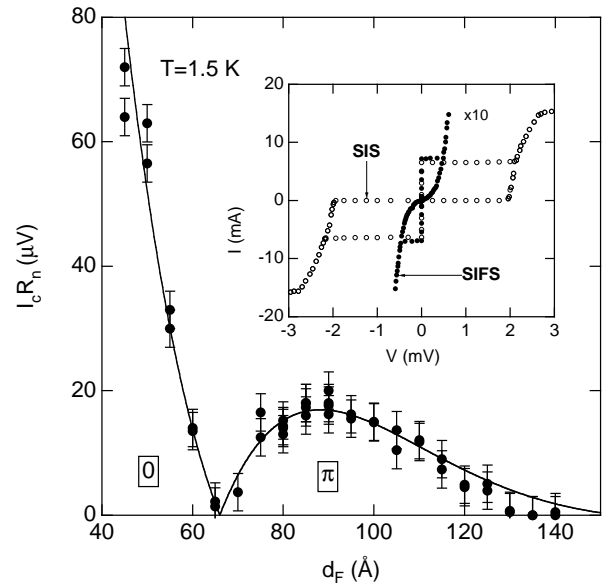}
\caption{Josephson coupling as a function of thickness of the PdNi layer (full circles). The critical current cancels
out at $d_{F} \simeq$ 65 {\AA} indicating the transition from "$0$" to "$\pi$"-coupling. The full line is the best fit
obtained from the theory (eqn.(1)) as described in the text. Inset shows typical I-V characteristics of two junctions
with (full circles), and without (empty circles) PdNi layer.}\label{fig2}
\end{figure}

The $Nb/Al/Al_{2}O_{3}/Nb$ junctions without a ferromagnetic layer showed high quality tunneling. At low temperature
the Nb/Al bilayer is a homogeneous superconductor. The I-V characteristic is hysteretic as expected for SIS junctions
\cite{bar82}. The subgap conductance at 1.5 K is only  $10^{-4}$ of the high energy conductance corresponding to
negligible current leakage as shown in the inset of Fig.\ \ref{fig2}. The $I_{c}R_{n}$ product is 1.25 mV with junction
to junction fluctuations of at most 15 $\%$. The inset of Fig.\ \ref{fig2} also shows the I-V characteristics of a
$\text{Nb/Al/Al}_{2} \text{O}_{3}\text{/PdNi/Nb}$ junction with a thin layer of PdNi ($d_{F}$ =50 {\AA}). The critical
current is strongly reduced while the subgap conductance is enhanced. Both effects result from the suppression of the
order parameter in F by the exchange field. The superconducting correlations are so efficiently reduced at the
$\text{Al/Al}_{2} \text{O}_{3}\text{/PdNi}$ interface \cite{kon01} that the dissipative branch of the I-V curve
reflects the energy dependence of the density of states in the Nb/Al bilayer.

The Josephson coupling, $I_{c}R_{n}$, as a function of the thickness of the ferromagnetic layer is presented in the
main body of Fig.\ \ref{fig2}. We chose this representation to eliminate fluctuations in the critical current that are
simply due to fluctuations in the junction resistance. For each value of $d_{F}$, the $I_{c}R_{n}$  product of two
different junctions is reported, with typical sample to sample variations of less than 10 $\%$, illustrating the
reproducibility of the junction fabrication technique. The Josephson coupling is of the order of 10-20 $\mu V$, i.e.
100 times lower that that measured on junctions without PdNi. The sign of $I_{c}$ cannot be determined from the I-V
characteristics and the transition from $"0"$ to $"\pi"$ coupling is revealed as a zero of the $I_{c}R_{n}$ product.
The latter occurs at  $d_{F} \simeq$ 65 {\AA}, consistent with our recent measurements of the density of states in F
\cite{kon01}. Fig.\ \ref{fig2} also shows the best fit to the theory in which the $I_{c}R_{n}$ product is obtained by
integrating the spectral current density found solving the Usadel equations for a SIFS junction (see Fig.\ \ref{fig1})
:

\begin{equation}\label{eq:couplageSIFIS}
I_{c} {R_{n}} = \frac{\pi}{2} \frac{\Delta}{e} \frac{1}{\gamma_{b}} \frac{\cos\frac{d_{F}}{\xi_{F}}\cosh\frac{d_{F}}{\xi_{F}}+\sin\frac{d_{F}}{\xi_{F}}\sinh\frac{d_{F}}{\xi_{F}}}{\cos^{2}\frac{d_{F}}{\xi_{F}}\cosh^{2}\frac{d_{F}}{\xi_{F}}+\sin^{2}\frac{d_{F}}{\xi_{F}}\sinh^{2}\frac{d_{F}}{\xi_{F}}}
\end{equation}

where $\gamma_{b}$ is the PdNi/Nb interface resistance. The Josephson coupling decreases linearly with $\gamma_{b}$ as
do the number of Andreev reflections at the S/F interface \cite{note2}. The fitting parameters are $\gamma_{b}=5.3$ and
$\xi_{F}$= 28{\AA}. The interface resistance is the same as that deduced from the thickness dependence of the zero
energy density of states \cite{kon01}, while the coherence length in PdNi is smaller possibly because of the few per
cent increase in the Ni concentration. The corresponding exchange energy is $E_{ex} = 35meV$, taking
$v_{F}=2\times10^{7} cm/s$ and the mean free path in the PdNi equal to $d_{F}$ as found by resistivity measurements on
bare PdNi thin films.

\begin{figure}
\includegraphics[width=8.0cm,keepaspectratio=true]{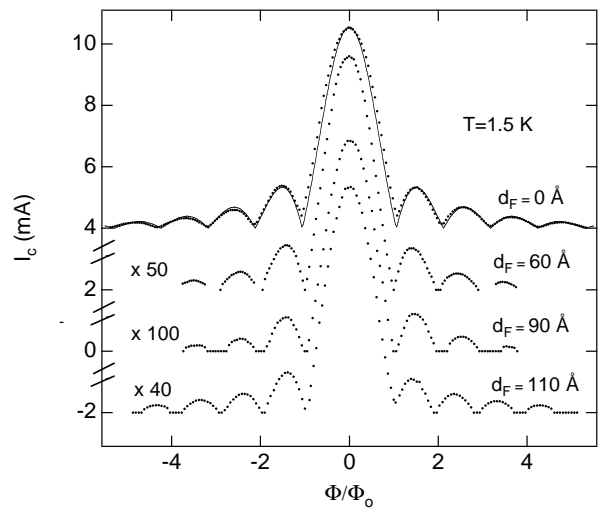}
\caption{Critical current as a function of magnetic flux for four different junctions, three with and one without PdNi.
The data are shifted vertically for clarity. The full line is the Fraunhofer pattern expected for a square junction
with homogeneous current density and a misalignment of $7^{\circ}$ between the field and the junction edge. The
junctions with PdNi have about the same coupling. The diffraction patterns are not affected by the PdNi
magnetization.}\label{fig3}
\end{figure}

In  Fig.\ \ref{fig3} the critical current, $I_{c}$ as a function of the magnetic flux is presented for (i) a junction
without PdNi and (ii) three junctions with different thickness of PdNi but roughly the same $I_{c}R_{n}$ product. One
of the latter gives $"0"$ coupling (60 \AA) and two $"\pi"$ coupling ( 90 {\AA} and 110 {\AA}). The diffraction
patterns are those expected for a square junction with homogeneous current density \cite{bar82} provided a small
misorientation of $7^{\circ}$ between the field direction and one edge of the junction is included (see fit in Fig.\
\ref{fig3}). The critical current cancels out when an integer of flux quanta settles into the junction. As the field
penetrates in a volume whose cross-section is given by $w(2\lambda+d_{F}$), where $\lambda$ is the penetration depth
and $w$=1 mm the junction width, an estimate of $\lambda$ for H=0.2 Gauss corresponding to the first quantum flux, is
$\lambda$=500 \AA, consistent with the value expected for Nb thin films \cite{bar82}. Importantly, the maximum critical
current occurs at zero applied field independently of the ferromagnetic layer thickness. This shows that the Josephson
coupling presented in Fig.\ \ref{fig2} is indeed the maximum coupling as assumed. It also indicates that the field
produced by the thin PdNi film falls off rapidly in the superconductor when one moves away from the S/F interface.
Therefore no orbital dephasing is generated in S by the ferromagnetic layer which acts only through the spin degree of
freedom as expected in the Pauli limit \cite{clo62}.

\begin{figure}
\includegraphics[width=8.0cm,keepaspectratio=true]{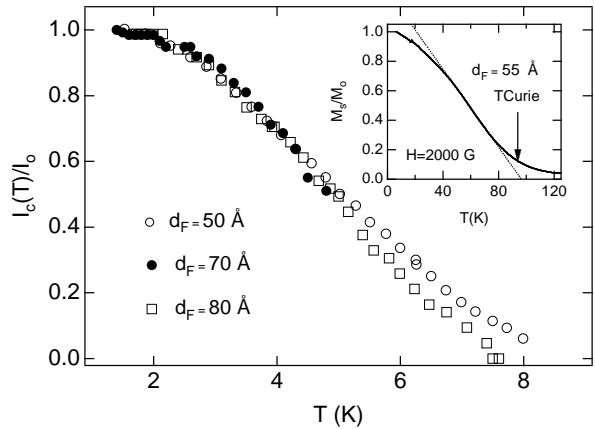}
\caption{ Normalized temperature dependence of the critical current for three different PdNi thicknesses. Note linear
dependence expected for Superconductor-Insulator-Normal-Superconductor tunnel junction \protect\onlinecite{row76}.
Inset shows normalized saturation magnetization as a function of temperature, measured via anomalous Hall effect for a
bare PdNi thin film.}\label{fig4}
\end{figure}

The temperature dependence of the critical current is found by integrating the spectral current density, $Q(\epsilon)$
multiplied by the distribution function \cite{yip98}. Far from $d_{F} \simeq 3\pi /4 \xi_{F}$, where the critical
current cancels out,  $I_{c}$ is mainly determined by the temperature dependence of the density of states on each side
of the junction. As the energy dependence of the density of states in the Al and PdNi is not exactly the BCS density of
states, the temperature dependence of $I_{c}$  shows a more linear behavior \cite{row76} than that originally found by
Ambegaokar-Baratoff \cite{amb63} for two bulk superconductors separated by a tunnel barrier. This linear behavior is
shown in Fig.\ \ref{fig4} where the temperature dependence of the normalized critical current for three junctions with
different PdNi thickness  is reported ($d_{F}$ = {50\AA} for "$0$" coupling, $d_{F}$ = {80\AA} for "$\pi$" coupling and
$d_{F}$ = {70\AA} near the transition ). The critical current of the junction with a {70\AA} thick PdNi layer is too
small to be measured at T $>$ 5K.

When $d_{F} \simeq 3\pi /4 \xi_{F}$, where the critical current cancels, the spectral current density $Q (\epsilon)$,
changes sign as a function of energy. Thus, changing the occupation of the ABS by raising the temperature allows to
shift the spectral  weight from a positive to a negative current density or equivalently from forward to back-ward ABS.
This results in a temperature induced "0-junction"-"$\pi$-junction" transition as recently observed in SFS junctions
\cite{rya01}. Such a transition only occurs for junctions very close to the critical point and it requires a
$I_{c}R_{n}$ product lower than 1$\mu$V. Because of their much smaller junction resistance, SFS junctions are more
suitable than SIFS to investigate small Josephson coupling. In fact, the minimum $I_{c}R_{n}$ product measurable in our
junctions is as low as a few $\mu$V and hence too large to reveal a sign change. As a final check we measured the
temperature dependence of the saturation magnetization of bare PdNi thin films ($d_{F}$=40-150 {\AA}) via the anomalous
Hall effect \cite{ber80}. As shown in the inset of Fig.\ \ref{fig4} for $d_{F}$= {55 \AA}  the normalized saturation
magnetization , $M_{s}/M_{0}$, which is proportional to the Hall signal at H=2000 Gauss, shows very small variations at
low temperature (T $<$ 10 K) corresponding to a few per cent change in $M_{s}$. Thus the long range magnetic order at
low temperature is well defined and the small change in $M_{s}$ has negligible effects on the critical current
temperature dependence.

In summary, we have shown that the critical current of SIFS Josephson junction is an oscillating function of the
thickness of the ferromagnetic layer. The zero of the critical current obtained for $d_{F} \simeq 3\pi /4 \xi_{F}$
indicates a change of sign of $I_{c}$, leading to "$\pi$-coupling". This is a consequence of the superconducting order
parameter oscillations induced in the ferromagnet by the proximity effect as recently observed by tunneling
spectroscopy \cite{kon01}. Such ferromagnetic-based "$\pi$-junctions" may represent a new element in the architecture
of novel quantum devices as recently suggested \cite{iof99}.

We are indebted to I. Balladi\'e and A. Buzdin for an illuminating conversation at the beginning of this work. We
acknowledge fruitful discussions with W. Guichard, P. Gandit, B. Leridon, J.C. Villegier, P. Feautrier, M. Salez. We
thank H. Bernas for a critical reading of the manuscript.


\begin{thebibliography}{99}
\bibitem{note2} Note that rough interfaces as well as a reduced probability of Andreev reflections on both sides of the ferromagnetic layer, randomize the quasiparticule trajectory in F as required in the dirty limit.
\end{thebibliography}
\end{document}